\documentclass[12pt,notitlepage,a4paper]{article}

\usepackage[utf8]{inputenc}
\usepackage{graphicx}
\usepackage{hyperref}
\usepackage[round,comma,authoryear]{natbib}
\usepackage[english]{babel}
\usepackage{amsmath}
\usepackage{amssymb}
\usepackage{lscape}
\usepackage{setspace}
\usepackage{CJKutf8}
\usepackage{tabularx,booktabs,multirow}
\usepackage[nameinlink,capitalize]{cleveref}
\usepackage{arydshln}

\usepackage{enumerate}
\usepackage[dvipsnames]{xcolor}
\usepackage{appendix}
\usepackage{authblk}
\usepackage[multiple]{footmisc}
\usepackage{tikz-network}
\usepackage{sidecap}
\usepackage{geometry}
\usepackage{titlesec}
\titlespacing*{\section}{0mm}{5mm}{2mm}
\titlespacing*{\subsection}{0mm}{3mm}{2mm}
\titlespacing*{\subsubsection}{0mm}{3mm}{2mm}

\definecolor{jpcol}{RGB}{231,138,195}
\definecolor{epcol}{RGB}{102,194,165}
\definecolor{uscol}{RGB}{141,160,203}

\usepackage[normalem]{ulem} 

\title{Multilayer patent citation networks: A comprehensive analytical framework for studying explicit technological relationships}
\author[1]{Kyle Higham\thanks{higham@iir.hit-u.ac.jp}}
\author[2]{Martina Contisciani\thanks{martina.contisciani@tuebingen.mpg.de}}
\author[2]{Caterina De Bacco\thanks{caterina.debacco@tuebingen.mpg.de}}
\affil[1]{Institute of Innovation Research, Hitotsubashi University, Japan}
\affil[2]{Max Planck Institute for Intelligent Systems, Cyber Valley, Tübingen, Germany}

\date{\vspace{-1cm}}

\begin{document}
\pagenumbering{gobble}
\maketitle

\begin{abstract}
    The use of patent citation networks as research tools is becoming increasingly commonplace in the field of innovation studies. However, these networks rarely consider the contexts in which these citations are generated and are generally restricted to a single jurisdiction. Here, we propose and explore the use of a multilayer network framework that can naturally incorporate citation metadata and stretch across jurisdictions, allowing for a complete view of the global technological landscape that is accessible through patent data. Taking a conservative approach that links citation network layers through triadic patent families, we first observe that these layers contain complementary, rather than redundant, information about technological relationships. To probe the nature of this complementarity, we extract network communities from both the multilayer network and analogous single-layer networks, then directly compare their technological composition with established technological similarity networks. We find that while technologies are more splintered across communities in the multilayer case, the extracted communities match much more closely the established networks. We conclude that by capturing citation context, a multilayer representation of patent citation networks is, conceptually and empirically, better able to capture the significant nuance that exists in real technological relationships when compared to traditional, single-layer approaches. We suggest future avenues of research that take advantage of novel computational tools designed for use with multilayer networks.
\end{abstract}

\clearpage
\pagenumbering{arabic}
\section{Introduction}

Patent citations have found successful application in a wide swathe of contexts, from understanding knowledge spillovers~\citep{jaffe1993geographic,sorenson2006complexity,jaffe2017patent,berkes2021geography} to the characterization of technological change~\citep{fleming2001recombinant,choi2009monitoring,huenteler2016technology}. The vast majority of this research is conducted using only patent data from a single jurisdiction and often ignores important citation context. However, as innovation and patent filings become increasingly global endeavours~\citep{fink2016exploring,danguy2017globalization}, there are many situations where it is important to think of `the patent system' as a set of quasi-coordinated processes operating across jurisdictional boundaries~\citep{petit2021patent}.

This coordination is desirable because the same invention can be patented in multiple jurisdictions; there are clear efficiency gains to be made if information discovered or produced during the patent prosecution process can be shared between jurisdictions~\citep{chun2011patent}. Patent families arise because these related applications, which simultaneously progress through multiple patent offices, are legally linked through their first filing. The Paris Convention\footnote{Paris Convention for the Protection of Industrial Property (1883).} allows applicants to apply in multiple jurisdictions and claim the filing date of the first application, known as the priority date, as the effective filing date for subsequent applications (provided these occur within 12 months of the priority date). Information sharing between offices creates, by design, some redundancy in the information generated for family members across offices, but not enough that a complete picture can be pieced together from a single jurisdiction's data. The existence of patent families provides the opportunity to form the most complete set of information about a particular invention that can be obtained from patent data and allows us to link metadata across jurisdictional boundaries~\citep{nakamura2015effect}. In this work, we demonstrate the utility of these linkages in the context of patent citation networks. 

The family-level view would suggest that only using data from a single jurisdiction leaves a lot of potentially relevant information unexamined~\citep{bakker2016patent}. In the more and more common scenario where multiple family members exist across multiple jurisdictions, citations will often only be made to one family member.\footnote{Search reports will often list equivalents of the prior art that is cited, however, this additional information is not explicitly included in the associated data sets.} As such, the citation network that is obtained from any single jurisdiction necessarily represents a subset of the complete network for the set of inventions under examination. While information sharing between offices will increase the amount of overlap between these networks, it does not make family-level analyses redundant, for two reasons. First, the amount of information sharing, and the modes for doing so, between patent offices has changed significantly in recent years.\footnote{See, e.g., \url{https://www.wipo.int/case/en/}.} In particular, advances in information technology allow patent offices to coordinate much more effectively than they did 20 years ago. Yet, some patents filed 20 years ago are only expiring now, so these patents can still be important sources of information when studying contemporary innovation. Second, many patents are \textit{not} filed in multiple offices, and some applicants only select a few strategically important jurisdictions where they would like to protect their intellectual property. That is, the nodes and links in the citation networks of each jurisdiction are unique, and so each network contains a huge amount of potentially pertinent information that is unique to that jurisdiction. In the empirical sections of this work, we take a very conservative approach and only consider nodes that are shared across jurisdictions, as described in detail in \Cref{sec:data}.

Even for shared nodes (defined here as patents granted in multiple jurisdictions), however, the sets and types of citations made by each patent office can differ greatly, as shown in \Cref{fig:multilayer}. The primary reason for this disagreement is that different jurisdictions abide by different legal guidelines that describe when and how citations should be made. These sets of guidelines are not without strong similarities, however, and a careful reading offers pathways towards sensible aggregation or comparison of these sets of citations~\citep{higham2022patent}. This has become particularly feasible in recent times as more and more offices now provide metadata about citation context, such as whether the cited patent was so similar to the application as to render the latter unpatentable, or whether the cited patent was added to simply define the state of the art. A secondary reason for disagreement between jurisdictions is, in fact, a commonality: examiners in all jurisdictions are humans with limited time to examine any particular patent~\citep[see, e.g.,][]{frakes2017time}. Often, it is simply not possible to find every relevant piece of prior art, particularly when language barriers are taken into consideration. Indeed, in combination with simple differences of opinion, this limitation means it is unlikely that two examiners in the same patent office would find exactly the same set of prior art~\citep{wada2016obstacles}. Therefore, using family-level information gives us the search result of more examiner-hours as well as the multiple opinions of what should be considered relevant prior art. 

As citations made by different offices are made according to different sets of guidelines, treating these citations as equally informational may lead to misleading results. Indeed, some suggest that citations of the \textit{same} type have become less informative over time~\citep{kuhn2020patent}. It is therefore important to aggregate family-level information sensibly. We propose that a multilayer network framework provides a natural representation of the patent citation network that readily incorporates differences in citation type. After all, multiple networks anchored by common nodes is the very definition of a multilayer network~\citep{de2013mathematical,kivela2014multilayer,porter2018multilayer}.

Within the multilayer framework, described in more detail for our chosen context in \Cref{sec:multilayer}, each layer of the network represents a single link type, each node represents a patent family (which may exist in multiple layers), and each link represents a citation between families (of the type defined by the layer). As such, the global patent citation network is an inherently multilayer system; no abstraction is required. Further, this framework is particularly flexible. For example, layers can represent jurisdictions, and links within each jurisdictional layer can represent the citations found on the front page(s) of the family member(s) granted by that jurisdiction. From this point, it is possible to layer as many jurisdictions as desired onto the network, provided there are family linkages existing between the layers. It is also possible to split these layers further, according to citation metadata that inform us of the reason for, or source of, a particular citation. This flexibility is particularly valuable when certain types of citations are irrelevant, or may even be considered pure noise, with respect to a particular research question. For example, one studying knowledge flow within a multilayer framework may not wish to consider citations discovered by the examiner, and may even want to add an additional layer for citations found in the patent specification~\citep{verluise2020missing}.

However, it is not clear, a priori, whether a multilayer framework adds any information over and above that which can be found in `flattened' family citation networks wherein citation context is disregarded and only link existence is examined~\citep{nakamura2015effect}. Thus, in order for the multilayer framework to be feasible as a research tool, it is important to first demonstrate a significant gain in information content relative to the flattened, global family citation network, or even the more commonly-used jurisdictionally restricted citation networks. To this end, we explore the information content of the triadic patent family network, wherein all layers contain the same set of nodes. This set consists of families containing at least one member granted in each of the triadic patent offices: the United States Patent and Trademark Office (USPTO), the European Patent Office (EPO), and the Japan Patent Office (JPO). The triadic offices have historically granted the majority of patents globally and contain rich and accessible citation information. Specifics about the data used in this work can be found in \Cref{sec:data}.

In this work, we first construct a multilayer family-family triadic citation network, wherein layers can be separated by jurisdiction and citation context (such as whether the citation was added by an applicant or examiner). In practice, the appropriate set of contexts can be selected based on the use-case; in this work, the additional context we consider is whether a citation was likely to have been found by the examiner or by the applicant (which is not always explicit), as we expect the differing motivations for citation between these groups affect the nature of the technological relationships reflected by these citations. We then conduct an interdependence analysis to check for redundancy of information content between the layers, finding that significant complementary information exists between jurisdictions. A community detection procedure is then conducted on the multilayer network and two comparison networks: the flattened multilayer network containing the same set of links but without information about jurisdiction or citation context, and the US-only subset of the citation network, also flattened. The former comparison tests the role of citation context, while the latter is included as the most commonly used patent citation network in prior research. We observe, graphically, nuanced differences in inferred community structure between the multilayer network and the comparison networks.

To add colour to these differences, we examine the relationships between inferred community partitions and the technology classes of the families that comprise them. For the multilayer network communities and those of the two flattened comparison networks, we project the bipartite community-class network onto the class nodes and directly compare these projections with established class-class networks (co-classification and inter-class citation linkage) with known-node-correspondence methods. We are also able to directly measure the diversity of communities, and the spread of classes between communities to inform our interpretation of the direct network comparisons.

When compared to the other two networks, we find that the multilayer case produces communities that more closely reflect the known technological relationships implied by the established class-class networks, at both micro- and meso-scales. Further, while technological classes are more splintered across communities in the multilayer case, the internal diversity of communities is lower than the comparison networks once we account for the known technological similarity of classes. These results suggest that, even within our conservative empirical framework, citation context is an important source of information about the nature and importance of the particular technological relationships codified by citation linkages, and that examination of multilayer citation networks using novel computational techniques is an exciting and relevant avenue for future research.

The rest of the paper is structured as follows. \Cref{sec:methods} introduces both patent families and multilayer networks and discusses how the former naturally forms the latter in the context of citation networks. \Cref{sec:data} describes the data we use in this work, how this forms the multilayer networks and why specific subsets of families and citations are selected for analysis. \Cref{sec:results} describes the empirical procedures that we use to test and compare the information content of the multilayer citation network relative to single-layer networks and describes the results obtained. Lastly, \Cref{sec:conclusions} concludes and discusses the limitations and extensions of this research.

\section{Multilayer patent networks}\label{sec:methods}

\subsection{Patent Families}\label{sec:families}

The rights bestowed by patents are only enforceable in the jurisdiction in which the patent was granted. To obtain these rights in more than one jurisdiction, an applicant first files in a single jurisdiction (often their local patent office), starting the clock on the period during which they can file for the same invention in other jurisdictions. For the next 12 months, all subsequent filings can `claim priority' from this initial application and inherit the latter's filing date as its own for the purposes of examination (provided the same content is covered in the application).

There are two primary modes through which an invention can claim priority from an earlier application: the Paris Convention and the Patent Cooperation Treaty (PCT). The former lays down the guidelines for the treatment of foreign patent applications among the contracting parties, including the time limit on priority claims as described above. The latter, for our purposes here, is effectively an attempt to streamline and harmonise the process of patenting in multiple jurisdictions.\footnote{\url{https://www.wipo.int/pct/en/}.} This process does not result in a patent, but rather a preliminary prior art search report, and allows the applicant to nominate the jurisdictions to which they would like to apply for a patent without having to apply at each office separately. Priority can be claimed from a PCT filing, and PCT filings can themselves claim priority from an earlier filing at a local office.

After a patent application has reached a local office, the applicant may want to fine-tune their claims or even be asked to split the described invention into two separate patent applications.\footnote{See, e.g., Paris Convention for the Protection of Industrial Property (1883), Article 4G.} The inventor is not able to disclose new information during this process, and thus the claims made by the `new version' of the application must be contained within the scope of the initial disclosure. These subsequent filings may claim the priority date of the initial filings and are referred to as `continuing applications'.

Patent families, in general, link patents and applications through their priority filing. The resulting `family trees' can be complex and, as such, several types of families exist~\citep{martinez2010insight,martinez2011patent}. `Simple' patent families (as defined by the EPO for their DOCDB database) each consist of a set of patents and applications that are all linked to the same priority filing. This type of family is the one on which we focus in this work, and we will henceforth drop `simple'. As such, families can be made up of sets of documents from several jurisdictions, each of which may contain multiple documents. Other families may only consist of a single application in a single jurisdiction.

Families are the unit of analysis for the current work for two reasons. First, they generally align with what one usually thinks of as a single `invention'~\citep{martinez2011patent} and it is the relationships between inventions that we usually aim to capture with citation data. Second, they link inventions across jurisdictions, and therefore allow the alignment of jurisdiction-specific citation networks and, therefore, the introduction of the multilayer network as a potentially useful analytical, conceptual, and mathematical tool to study technological relationships.

From the perspective of data availability, detail, and volume, the obvious choice of data set for testing the utility of the multilayer framework are those patents granted by the three (historically) largest patent offices, also known as the triadic offices: the USPTO, EPO, and JPO. Further, we wish to take a particularly conservative empirical approach to these initial explorations of the multilayer citation network. To do this, we only consider patent families that have granted members in all layers of interest (`triadic families') and only consider citations among these families.\footnote{The applicants to these offices, however, may be based outside these jurisdictions.}

Theoretically, each office examining these triadic applications has access to the same information regarding prior art, and they share much of what they find with the other two offices, directly or indirectly~\citep{wada2020uspto,petit2021patent}. For this reason, granted members have all had the same opportunities to link to other (older) families in each layer, maximising potential redundancies between layers in the citation network. The exclusion of families that are not triadic, therefore, is why we think of this analysis as likely to produce very conservative results when compared to those that may be obtained for a network without such exclusions.

We also note triadic patent families are often used as a binary indication of a `high-quality' invention~\citep{de2009policy,tahmooresnejad2019capturing}; after all, the applicants thought it was worth the time and money to patent their invention in three of the largest markets in the world. By this logic, our multilayer network consists exclusively of `high-quality' patent families\footnote{Note that this is a very narrow view of patent quality. For a comprehensive discussion, refer to \cite{higham2021patent}.} and excludes much controversial subject matter that are not universally patentable~\citep{biddinger2000limiting}. 

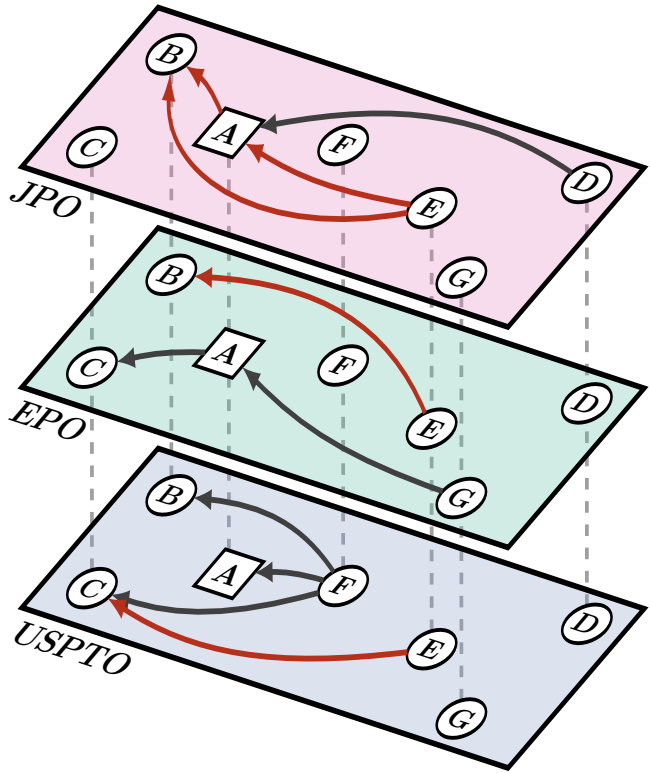
\begin{figure}[t]
    \caption{\textbf{Exemplar subset of the multilayer patent citation network.} A multilayer representation of a typical subset of the inter-family patent citation network we consider in this work. Nodes and links comprise the multilayer ego network of patent family \textbf{A}, the USPTO equivalent of which is ``Power source apparatus'' (US6819081B2), initially filed in January 2002 by Sanyo Electric Co., Ltd. at the JPO. Each layer represents the inter-family citations made by a different patent office, and red links are those used to justify a (non-final) rejection of the application that was examined in that layer. All data represented here is subject to the restrictions described in \Cref{sec:data} and is, therefore, an extremely simplified version of the complete ego network. Details of the families represented can be found in \Cref{app:ego_details}. \label{fig:multilayer}
     \vspace{3mm}}
\centering     
\resizebox{3.5in}{4in}{
    \begin{tikzpicture}[multilayer=3d]
   
    \begin{Layer}[layer=1]
        \Plane[x=0,y=0,width=2.4,height=7,opacity=.3,color=jpcol]
        \Vertex[x=2,y=1.2,label=\textbf{A},color=white, size=0.55, fontsize=\small, shape=rectangle]{A}
        \Vertex[x=0.6,y=1.9,label=\textbf{B},color=white,size=0.55, fontsize=\footnotesize]{B}        
        \Vertex[x=0.6,y=0.5,label=\textbf{C},color=white,size=0.55, fontsize=\footnotesize]{C}
        \Vertex[x=6.5,y=2.0,label=\textbf{D},color=white,size=0.55, fontsize=\footnotesize]{D}
        \Vertex[x=5,y=1.1,label=\textbf{E},color=white,size=0.55, fontsize=\footnotesize]{E}
        \Vertex[x=3.4,y=1.5,label=\textbf{F},color=white,size=0.55, fontsize=\footnotesize]{F}
        \Vertex[x=6,y=0.4,label=\textbf{G},color=white,size=0.55, fontsize=\footnotesize]{G}
                
        \Edge[bend=10,Direct=true,NotInBG,color=BrickRed](A)(B)      
        \Edge[bend=23,Direct=true,NotInBG](D)(A)              
        \Edge[bend=-10,Direct=true,NotInBG,color=BrickRed](E)(A)              
        \Edge[bend=-43,Direct=true,NotInBG,color=BrickRed](E)(B) 
        \node at (0,0)[below right]{\textbf{JPO}};
    \end{Layer}    
    \begin{Layer}[layer=2]
        \Plane[x=0,y=0,width=2.4,height=7,opacity=.3,color=epcol]
        \Vertex[x=2,y=1.2,label=\textbf{A},color=white,size=0.55, fontsize=\small, shape=rectangle]{H}
        \Vertex[x=0.6,y=1.9,label=\textbf{B},color=white,size=0.55, fontsize=\footnotesize]{I}        
        \Vertex[x=0.6,y=0.5,label=\textbf{C},color=white,size=0.55, fontsize=\footnotesize]{J}
        \Vertex[x=6.5,y=2.0,label=\textbf{D},color=white,size=0.55, fontsize=\footnotesize]{K}
        \Vertex[x=5,y=1.1,label=\textbf{E},color=white,size=0.55, fontsize=\footnotesize]{L}
        \Vertex[x=3.4,y=1.5,label=\textbf{F},color=white,size=0.55, fontsize=\footnotesize]{M}
        \Vertex[x=6,y=0.4,label=\textbf{G},color=white,size=0.55, fontsize=\footnotesize]{N}
        
        \Edge[bend=25,Direct=true,NotInBG,color=BrickRed](L)(I)
        \Edge[bend=-10,Direct=true,NotInBG](N)(H)  
        \Edge[bend=10,Direct=true,NotInBG](H)(J)
        
        \Edge[style=dashed,opacity=0.5](A)(H)
        \Edge[style=dashed,opacity=0.5](B)(I)
        \Edge[style=dashed,opacity=0.5](C)(J)
        \Edge[style=dashed,opacity=0.5](D)(K)
        \Edge[style=dashed,opacity=0.5](E)(L)
        \Edge[style=dashed,opacity=0.5](F)(M)
        \Edge[style=dashed,opacity=0.5](G)(N)

        \node at (0,0)[below right]{\textbf{EPO}};
    \end{Layer} 
    \begin{Layer}[layer=3]
        \Plane[x=0,y=0,width=2.4,height=7,opacity=.3,color=uscol]
        \Vertex[x=2,y=1.2,label=\textbf{A},color=white,size=0.55, fontsize=\small, shape=rectangle]{O}
        \Vertex[x=0.6,y=1.9,label=\textbf{B},color=white,size=0.55, fontsize=\footnotesize]{P}        
        \Vertex[x=0.6,y=0.5,label=\textbf{C},color=white,size=0.55, fontsize=\footnotesize]{Q}
        \Vertex[x=6.5,y=2.0,label=\textbf{D},color=white,size=0.55, fontsize=\footnotesize]{R}
        \Vertex[x=5,y=1.1,label=\textbf{E},color=white,size=0.55, fontsize=\footnotesize]{S}
        \Vertex[x=3.4,y=1.5,label=\textbf{F},color=white,size=0.55, fontsize=\footnotesize]{T}
        \Vertex[x=6,y=0.4,label=\textbf{G},color=white,size=0.55, fontsize=\footnotesize]{U}

        \Edge[bend=20,Direct=true,NotInBG](T)(P)
        \Edge[bend=10,Direct=true,NotInBG](T)(O)
        \Edge[bend=-20,Direct=true,NotInBG](T)(Q)
        \Edge[bend=-20,Direct=true,NotInBG,color=BrickRed](S)(Q)
        
        \Edge[style=dashed,opacity=0.5](O)(H)
        \Edge[style=dashed,opacity=0.5](P)(I)
        \Edge[style=dashed,opacity=0.5](Q)(J)
        \Edge[style=dashed,opacity=0.5](R)(K)
        \Edge[style=dashed,opacity=0.5](S)(L)
        \Edge[style=dashed,opacity=0.5](T)(M)
        \Edge[style=dashed,opacity=0.5](U)(N)
   
        \node at (0,0)[below right]{\textbf{USPTO}};
    \end{Layer}   
    \end{tikzpicture}
    }
\end{figure}

A simplified diagram of a multilayer network of citations between triadic families is shown in \Cref{fig:multilayer}, with full details of the families included in this example described in \Cref{app:ego_details}. Note that the multilayer network that we analyse in this paper treats sub-layers, such as whether a citation has been used in a rejection decision (shown in red in \Cref{fig:multilayer}), as distinct layers. This results in seven layers in total, as the EPO also provides information about whether a citation originated from the international search report (conducted outside the EPO) or the local search report.

\subsection{Multilayer networks}\label{sec:multilayer}

Multilayer networks have received particular attention in the past decade~\citep{de2013mathematical, kivela2014multilayer,boccaletti2014structure,cimini2019statistical}, and the development of mathematical and computational tools for their analysis, as well as their timely application, remains a very active field of research across many domains~\citep{gallotti2016lost,vaiana2020multilayer,harvey2020network,yuvaraj2021topological,van2021framework}. In this work, we not only suggest that patent citation networks are naturally multilayered, but aim to introduce the multilayer framework to the innovation studies community to promote the timely application of novel computational tools that are currently being developed.

To date, the vast majority of the studies that explicitly place patent citation data into a network setting use a single-layer framework~\citep{von2005inventive,valverde2007topology,clough2015transitive,nakamura2015effect,funk2017dynamic,wu2019large,higham2019ex,mariani2019early}. That is, there is only one type of link (i.e., a citation) between nodes in the network. This approach often makes practical sense, such as when one lacks citation metadata that may be used to distinguish or `colour' the links, or if only one link type is of interest. However, a multilayer network framework is able to naturally incorporate citation metadata, if it exists, into the network structure.\footnote{In a related work, also using triadic patents, \cite{morrison2014border} use a multiplex PageRank to assess the centrality of technology classes where layers are defined by inventor location. However, citation source and context are not considered.} As an analogy, let us consider the public transport network of a large city containing several different forms of transport, each with its own network of routes and stations. There are usually many points of overlap between these network layers to allow passengers to transfer between modes of public transport, such as a bus stop at a train station. These transfer points link the different network layers together. From both mathematical and computational perspectives, this kind of network is fundamentally different from single-layered networks, particularly when the different layers are defined by links with very different properties~\citep{aleta2017multilayer,ibrahim2021optimal}. In the public transport context, these properties can be straightforward, such as speed, price, comfort, or environmental harm, or more computationally complex, such as sensitivity to link removal and amenability to rerouting~\citep{de2014navigability}.

In the domain of patent citation networks, each jurisdiction has a set of applications and patents that each contain a set of citations made to other documents. Each of those citations comes with context~\citep{higham2022patent}. This context can be whether the prior art was discovered by the examiner, the justification for its addition to the document, the relationship between the citing and citing firms, or any other citation metadata that may be obtained or constructed. For many research questions that rely on information derived from the citation network, this information is important to retain, just as it is important to know whether two nodes in a transport network are connected by a bus, an airplane, or a ferry. 

At the same time, every patent is part of a family (even if there is only one member). When families contain members filed in multiple jurisdictions, the citation networks associated with each jurisdiction can be linked, just as a bus may stop at a train station, or a train may stop at an airport. Of course, patent applicants are under no obligation to file for a patent on the same invention in multiple jurisdictions. That is, a node (patent family) may not exist in all layers of the network. Not every bus stop is a train station, nor vice versa. The full patent citation network is a true `multilayer' network in this sense. In this work, however, we focus on the subset of nodes that exist across all three layers of interest (the triadic offices). The justifications for this choice are discussed in \Cref{sec:data}. The network we define in this work, therefore, is a special case of a multilayer network wherein the layers are node-aligned~\citep{kivela2014multilayer}. Extensions of this work to a more general multilayer framework are discussed in \Cref{sec:conclusions}.

Multilayer networks share many characteristics of interest that are found in single-layer networks; indeed, much of the early research on multilayer networks involved adapting concepts from single-layer networks to this new framework~\citep{berlingerio2011foundations,brodka2012analysis,de2013mathematical,battiston2014structural}. For our purposes, in order to demonstrate the utility of the multilayer framework, it is necessary to compare the network properties derived in this setting to those obtained from the equivalent, flattened single-layer network, wherein citation metadata is ignored (partially or wholly). 

The domain within which we choose to explore differences between the multilayer and single-layer frameworks, in the patent citation context, is community detection. The natural grouping of nodes is one of the characteristic features of real-world networks and plays a significant role in describing the structure of the network at scales between node-level and global-level network statistics~\citep{wasserman1994social,newman2004finding,fortunato2010community}. Often, innovation researchers are interested in the composition of, and interaction between, close-knit groups of meso-scale objects such as groupings of similar technologies~\citep{lee2015predicting,alstott2017mapping,balland2017geography,yan2017measuring,mejia2020emerging}, and the application of community detection to the multilayer citation network leaves room for direct comparison between our results and these objects that we usually work with. Lastly, community detection can be applied to both multilayer and single-layer networks, which will allow for comparisons between the resultant communities.

\subsection{Data}\label{sec:data}

The multilayer citation network we construct is generated by citations made by triadic patents and only includes those made to and by triadic families. For the purposes of the current work, \textit{triadic patents} are patents granted by one of the triadic offices that have family members, or equivalents, granted by the other two triadic offices. \textit{Triadic families}, on the other hand, will refer to the full set of documents belonging to a family that contains triadic patents.\footnote{Note that this definition is slightly different to that used in previous work, notably \cite{dernis2004triadic}. Until the year 2000, applications to the USPTO were not published, so it was generally impossible to know whether equivalents were filed in all jurisdictions. This led to a slightly awkward definition (families with equivalents granted by USPTO and applied to EPO and JPO) that was in wide use until sufficient time had passed for USPTO application data to accumulate. A common definition in use currently is those families with equivalents filed at the triadic offices; however, as US applications do not list citations, we restrict this definition further to require a grant at each office.} These sets include both applications and patents and may be filed at or granted by offices outside the three triadic offices (provided that they are within a family containing triadic patents). 

There are several reasons for choosing this subset of nodes and links to define our network, beyond the aforementioned desire to be conservative in our empirical design. The first is that we require well-defined layers. By restricting the citing patents to those granted by the triadic offices, the links (and, therefore, network layers) are defined by the citation context (e.g., the jurisdiction where it was made and the reason it was added), which isn't available for many offices. Second, restricting the cited families to those that are also triadic means that there are no cross-layer citations, which significantly simplifies the network from a mathematical perspective. For example, a US triadic patent citing a pre-grant publication that was only filed at the Japan Patent Office would be a cross-layer citation, as the latter node does not exist in the US layer. If, however, this Japanese publication was part of a triadic patent family, we can `redirect' this US-originating citation to the US-granted family member, as this patent covers the same technical content, and the citation can remain within the US citation network layer where it was generated. Third, all triadic offices provide detailed citation data. There is no theoretical reason why citation network layers associated with other countries cannot be added if the data exists, but we deemed the triadic offices to be the best starting point to demonstrate the use of the multilayer framework due to their existing popularity among both applicants and researchers. 

\begin{figure}[t]
    \caption{\textbf{Family priority dates.} A histogram of the priority dates of the triadic families considered in this work, subject to the restrictions laid out in \cref{sec:data}. All families have their US member granted in 2001 or later, but the earliest filing date can be considerably earlier.}
    \vspace{3mm}
    \centering
    \includegraphics[width=\textwidth]{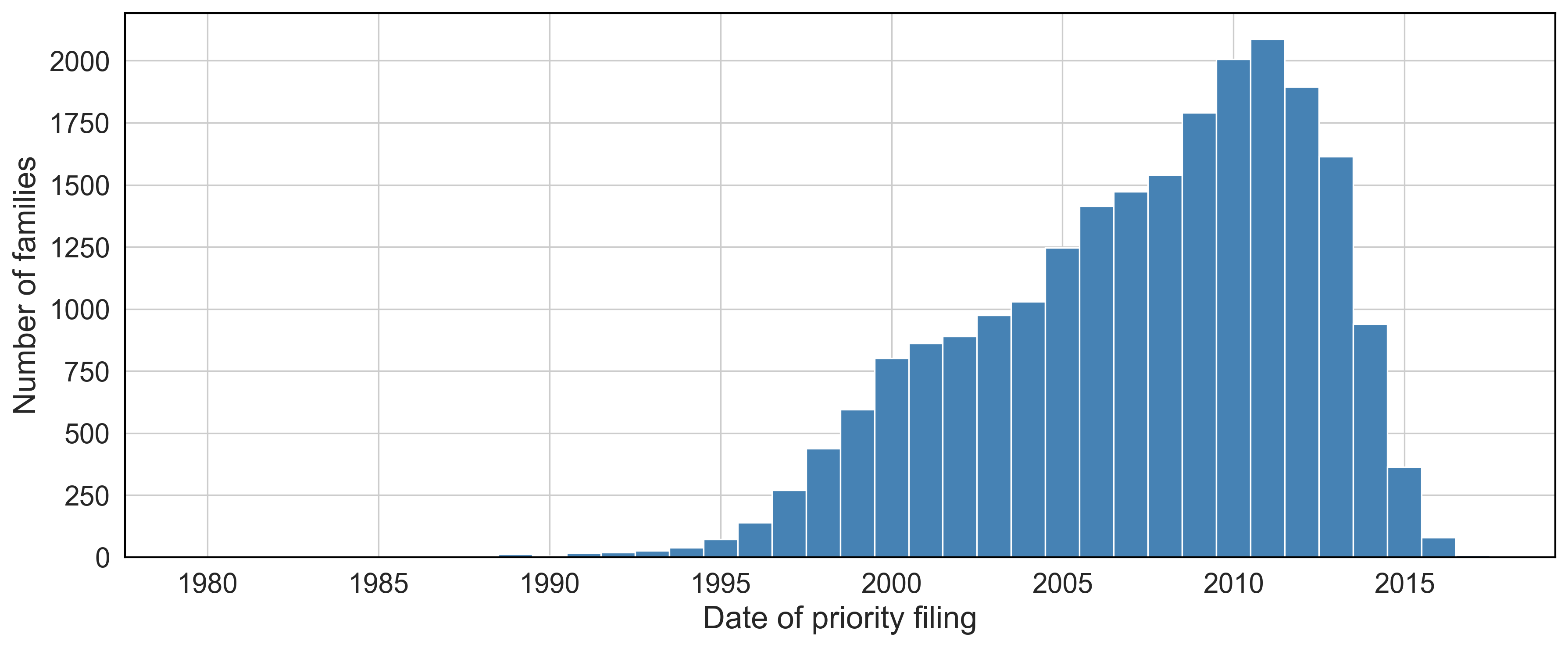}
    \label{fig:histogram}
\end{figure}

In this work, we also wish to demonstrate the importance of citation source and context. During the application and examination process, citations that reach the front page of the patent may be added by one of several parties for a variety of reasons. One problem inherent in this citation metadata is that different offices have different examination guidelines and legal frameworks that inform how prior art is cited~\citep{higham2022patent}. Further, the way that these differences manifest themselves in the metadata that researchers can access is not consistent across offices or, indeed, across time. For some of the analyses in this work, we broadly group citations at each office into two groups: those that were likely found by the examiner and those that were likely found by the applicant. While these groups are far from perfect,\footnote{This is particularly true for the JPO. However, there is suggestive evidence that applicant citations are more likely to be background art than art that could lead to a rejection of the application~\citep{okada2018making}.} we do so to illustrate the flexibility of the multilayer network approach---the citations that comprise each layer can be filtered based on the research purpose. This flexibility is discussed in more detail in \Cref{sec:conclusions}. One minor restriction that accompanies this approach is that we require citation metadata to exist for all citing patents. The USPTO only started to include this metadata for granted patents from the start of 2001, so the triadic families we consider in this work are those for which the first US grant was in 2001 or later. All families considered in this work have all of their triadic members granted before April 2020. A histogram of the priority dates of the families that comprise the networks we consider in this work is displayed in \cref{fig:histogram}.

Most of the data used in this work were obtained from Google Patent Public Datasets.\footnote{\url{https://tinyurl.com/googlepatentdata} (accessed 25/10/2021).} However, noting that, at the time of data collection, that data was not complete for citations between Japanese publications (notably, Japanese patents citing published Japanese applications), this data was supplemented by data supplied by the Intellectual Property Institute's Patent Database.\footnote{\url{www.iip.or.jp/e/patentdb/index.html} (accessed 25/10/2021).} We also make use of Cooperative Patent Classifications (CPCs); for consistency, we assign each family the classifications associated with their first US member, as determined by the USPTO. This data was obtained from PatentsView.\footnote{\url{https://patentsview.org/} (accessed 25/10/2021).}

To reduce the computational complexity associated with large networks, we prefer to work with a subset of the whole patent family network that nonetheless resembles the structure of the whole. Using a set of obviously technologically related families such as those in a specific technology class or filed by firms in a specific sector may not satisfy this requirement, given the known differences in citation patterns across fields~\citep{alcacer2009applicant,higham2017fame}. To remedy this, we choose the subset of patents assigned to CPC class Y02: ``technologies or applications for mitigation or adaptation against climate change.'' The Y02 class is always a secondary classification and can be added to patent families from a broad set of technologies, from those aimed at reducing drag on airplanes to those aimed at treating diseases whose impact may be exacerbated by climate change~\citep{veefkind2012new,havsvcivc2015measuring}. This class (and its subclasses) are commonly used as filters to study patented technological developments within specific domains related to both the mitigation of climate change, such as cleaner transport~\citep{aghion2016carbon,barbieri2016fuel} and energy production~\citep{sun2021dynamic,persoon2020science}, and our adaptation to the inevitable and wide-ranging environmental challenges we will face in the near future~\citep{dechezlepretre2020invention,hotte2021knowledge}. As such, we believe this technology class comprises a suitable microcosm within which we can effectively demonstrate the application of multilayer network methods to patent citation networks. 

The resulting data set consists of a well-defined set of citing families, their CPC classifications, the citations they make,\footnote{We exclude very rare citation types, such as those originating from third parties.} and the jurisdiction and context of each citation. A description of the layers considered in this work (which can be aggregated for specific empirical tests) can be found in \Cref{tab:sum_stats}.

\begin{table}[t]
\caption{ \textbf{Layer descriptions.} Descriptions of the layers considered in this work, alongside their abbreviations and the number of links found within them. All layers contain 22653 nodes, and there are a total of 63916 citations in the multilayer network (MULTI) that is comprised of the layers described in the first seven rows. The last two rows are single-layer networks obtained by flattening the two USPTO layers (US-AGG) and all seven layers (ALL-AGG), respectively.\label{tab:sum_stats} }
\vspace{3mm}
\centering
\renewcommand{\arraystretch}{1.2}

\begin{tabular}{cccp{5.5cm}c}
Layer & Citing party & Abbreviation & Description & Links          \\ \hline
USPTO & Examiner & US-EXM & Cited by examiner during patent prosecution &   15607         \\
USPTO & Applicant & US-APP & Cited by applicant through an Information Disclosure Statement and unused by examiner & 23145      \\
EPO & Applicant & EP-APP & Cited by applicant, in the patent text or otherwise &   4326         \\
EPO & Examiner & EP-ISR & Cited by examiner in an international search report &    5732       \\
EPO & Examiner & EP-SEA & Cited by examiner in an EPO search report &    5206     \\
JPO & Examiner & JP-REJ & Cited by examiner as justification for application rejection &      4612    \\
JPO & Examiner & JP-BCK & Cited by examiner as background information &       5288     \\
 \hline
USPTO & All & US-AGG & Cited by anyone (USPTO patents) &       38752    \\
All & All & ALL-AGG & Cited by anyone (all triadic patents)  &    63916     \\

\end{tabular}

\end{table}

\section{Methods and Results}\label{sec:results}

\subsection{Interdependence}\label{sec:interdependence}

Before a detailed examination into the \textit{kind} of information that may be extracted from the multilayer network that is not accessible when using a single layer, it is first important to assess whether there is new information in the multilayer network at all. That is, if there is a high level of redundancy between the information contained in each network layer, then the case for using a multilayer framework is weakened. At the same time, if the layers contain very different structural patterns, then a multilayer framework may not be ideal, and more informative results may be obtained if they are treated as individual single-layer networks instead.

One way of assessing these properties is by measuring the interdependence of each layer, or set of layers, relative to the information that can be found elsewhere in the network. Several measures of interdependence have been proposed in the past~\citep{parshani2011inter,morris2012transport,nicosia2013growing}, many of which take a random walk approach to the level of layer interdependence or `coupling' of layers in the network. In this work, at a high level, we are instead interested in the degree to which the information contained in one network layer can inform us about the information contained in another layer.

To this end, we employ the method introduced by \cite{de2017community} and described in detail for our case in \Cref{app:interdepedence_calc}. This method is a link prediction exercise, whereby a randomly-selected portion of the target layer or layers $\alpha$ have their link information removed and the remaining information in the network may be used to predict the existence of links. As a baseline, the remaining portion of the $\alpha$ is used as the training set, the receiver operating characteristic (ROC) curve is calculated, and the area under this curve (AUC) is computed. We can then introduce sets of other layers, $\beta$, into the training set, and compare results obtained by adding this information to those of the baseline. If the predictive power (as measured by the AUC) of this augmented set $\alpha+\beta$ is not significantly larger or smaller than the baseline predictive power, then $\beta$ does not contain useful information over and above that contained in $\alpha$. If, however, we note a significant increase in predictive power relative to the baseline, then $\beta$ contains complementary information that cannot be extracted from what remains of $\alpha$. 

Much information can be garnered from comparisons of the change in predictive power when $\alpha$ and $\beta$ are interchanged. For example, when the links in one layer are a subset of links in another, then we expect the change in predictive power to be asymmetric when we swap $\alpha$ and $\beta$ --- adding the subset to try to predict links in the full set will likely produce \textit{worse} results than if only the full set was used for training the model. The information in the subset is redundant and could even mislead the model. 

When two layers contain complementary information we would expect increases in predictive power regardless of the layer comprising the test set. This complementarity can arise in several ways, such as through similar community structure despite large differences in the specific links that produce these structures. A significance decrease, on the other hand, would indicate that $\beta$ contains information that is irrelevant for the prediction task and actually added noise; this could occur, for example, if the link generation mechanisms were independent of the node properties, or were driven by different node properties in different layers.

\begin{figure}[t]
    \caption{ \textbf{Layer interdependence.} The $x$-axis presents different target layers $\alpha$, and the $y$-axis shows the AUC obtained through 5-fold cross-validation for measuring layer interdependence. Orange results refer to the baseline AUC, where the algorithm is only given access to that target layer. Green and blue markers show the increase in the AUC for the $\alpha$ set when the algorithm is given access to the US-APP or the US-EXM, respectively. The red points refer to the AUC obtained by giving access to all other layers in the network. The results displayed are averages and standard deviations over the 5 folds.}
    \vspace{3mm}
    \centering
    \includegraphics[width=\textwidth]{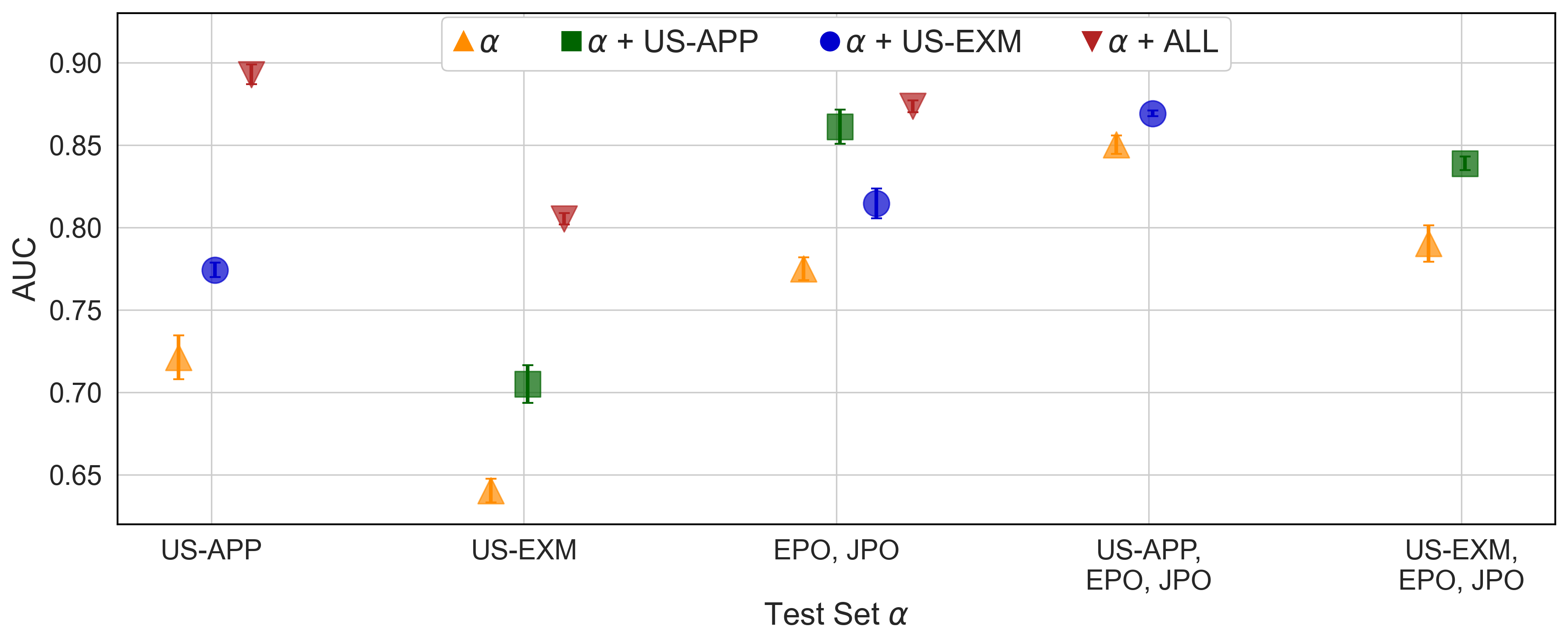}
    \label{fig:interdependence}
\end{figure}

\Cref{fig:interdependence} shows the results of the interdependence analysis for various $\alpha$ and $\beta$ sets in which we are interested. For graphical simplicity, we focus on the sublayers generated by the USPTO and the complete JPO and EPO layers (where the latter two always include all of their sublayers listed in \cref{tab:sum_stats}). This is done to demonstrate, compactly, the complementarity of information across jurisdictions as well as that of their sublayers, with the most commonly utilised sublayers in the literature (US applicant and examiner citations) as exemplars for the latter calculations. 

The results displayed in \Cref{fig:interdependence} show that adding more layers increases predictive power across all combinations of $\alpha$ and $\beta$ we considered. This outcome suggests that, while they differ by the amount of unique complementary information they contain, each layer nonetheless contains information that is not available in the other layers. Specifically, information about the missing values in $\alpha$ is more accurately predicted when layers that are not already in $\alpha$ are included in the training set, relative to the sole use of the information that remains in $\alpha$. This is to be expected, as examiners at each office conduct much of their prior art search independently. 

A prime example of complementarity is displayed by the US sublayers (US-APP and US-EXM). These layers are almost mutually exclusive,\footnote{Copying occasionally happens due to the recycling of citations for continuing patent applications.} but predictive power for links in one layer is significantly boosted when the other layer is added, regardless of which is the test set. That is, there is very little overlap in these layers, and yet one can be successfully used to predict the links in the other, likely due to the similarity of mesoscale network communities within each of these layers.

That some citation types add more information than others is also expected. After all, information sharing occurs regularly between offices~\citep{wada2020uspto}, and this process leads to the duplication of citations between specific layers. While this sharing happens increasingly through direct collaboration between offices examining equivalents,\footnote{\url{https://www.wipo.int/case/en/}.} most of the citations we consider here were made before these formal programs were launched. As such, for much of the time period we consider, the information `sharing' likely takes place indirectly, through applicants. For example, the EPO produces a search report for the applicant to consider before a substantial examination takes place. Under their duty of disclosure obligations at the USPTO, it is considered good practice to pass this information on to the USPTO if an equivalent is being examined there simultaneously (which will usually be the case for triadic patent families). This information is submitted via an information disclosure statement and the USPTO examiner then assesses the relevancy of the prior art that is listed on the search report. When it happens at all, only a small percentage of citations from the EPO search report will be used to justify rejection and be recorded as examiner citations, while the remainder will be recorded as applicant citations. As such, the EPO search report is a non-obvious mechanism through which citations are duplicated from EP-SEA citations to US-APP citations (and sometimes to US-EXM citations).

Similarly, while there is a knowledge disclosure obligation at the JPO, the incentives for complying are very weak relative to the USPTO~\citep{nakamura2016information}. However, applicants to the JPO often use in-text citations to make a case for patentability, and perhaps much more so than the typical applicant to the USPTO or EPO. As such, it is plausible that, for triadic patents, these citations are included in-text in other equivalent applications and are therefore easily accessible to examiners in all jurisdictions. If these citations are deemed relevant by multiple examiners, these citations might also appear to be duplicated across network layers.

\subsection{Community detection}\label{sec:community_detection}

Having found that the different layers likely contain complementary information, we now investigate the patterns extracted from a multilayer network approach and compare them with those extracted from single-layer networks that exclude citation context. Specifically, we wish to detect communities of triadic families that are similar in their citation patterns. These communities represent mesoscopic structural patterns contained in the networks that are not objectively or directly observed, but can be inferred from the data.

To this end, we apply a community detection algorithm to three networks: i) the (seven-layer) multilayer network containing the EPO, JPO, and USPTO layers (MULTI); ii) the network obtained by flattening all the layers in (i) into a weighted single-layer network and ignoring citation origin and context (ALL-AGG); iii) the weighted single-layer network obtained by flattening the USPTO examiner and applicant layers only (US-AGG). Each link is weighted by the sum of link weights across all layers we consider; that is, if the same family-family citation exists once in each of $n$ layers, then the link is assigned weight $n$. While rare, link weights greater than one can occur within sublayers; for example, when a divisional makes the same type of family-family citation as its parent, the link weight corresponding to this link will be two.

To perform the community detection task, we consider a probabilistic generative model that assigns a probability to a citation between two families that depends on the communities they belong to, as described in \cite{de2017community}. In our case we have access to relevant metadata about each triadic family, hence we consider the model of \cite{contisciani2020community}, MTCOV, that is also able to incorporate the office at which priority was filed (which is often \textit{not} a triadic office) as a node covariate to drive inference along with the network structural information. This covariate allows us to incorporate the home-bias of citations in early search reports~\citep{bacchiocchi2010international} and, to a lesser extent, industrial agglomeration patterns~\citep{asheim2005geography} (given the strong correlation between assignee location and priority office), to inform the inferred citation probability alongside explicit network structure. This model automatically balances the weight of the covariates' contribution in determining the communities. In all our experiments we find that node covariates are indeed significant, in that they allow us to better quantify the probability of certain citation patterns. The optimal number of communities in each case is extracted through a cross-validation procedure, see \Cref{app:MTCOV} for details. In addition to being able to incorporate a covariate that may inform network structure at scales beyond individual links, MTCOV is scalable to large networks, allows overlapping communities, and is open-source,\footnote{\url{https://github.com/mcontisc/MTCOV}} all of which are desirable features for the current work.

We chose ALL-AGG as a comparison because it contains all the same links as the multilayer network, and even accounts for link overlap among layers, but without context. As such, any differences in the extracted communities arise solely due to the addition of citation context, and the incorporation of this context into our network model. US-AGG is included in these comparisons as the most common citation network used in previous work. The USPTO also tends to make many more citations per patent, and so this single-jurisdiction layer is likely to be the most `complete', with respect to the links in the full triadic network.

\begin{figure}[t]
    \caption{\textbf{Community extraction.} This diagram shows the hard community membership partitions for MULTI. While inference was performed on the whole network, here we use a random sample of 2000 nodes and include any incidental links among these, for graphical clarity. The colouring shows the 15 communities found within the MULTI network. Node size is proportional to the number of outgoing and incoming citations, while node shapes denote the location of the assignee of each patent family.}
    \vspace{3mm}
    \centering
    \includegraphics[width=\textwidth]{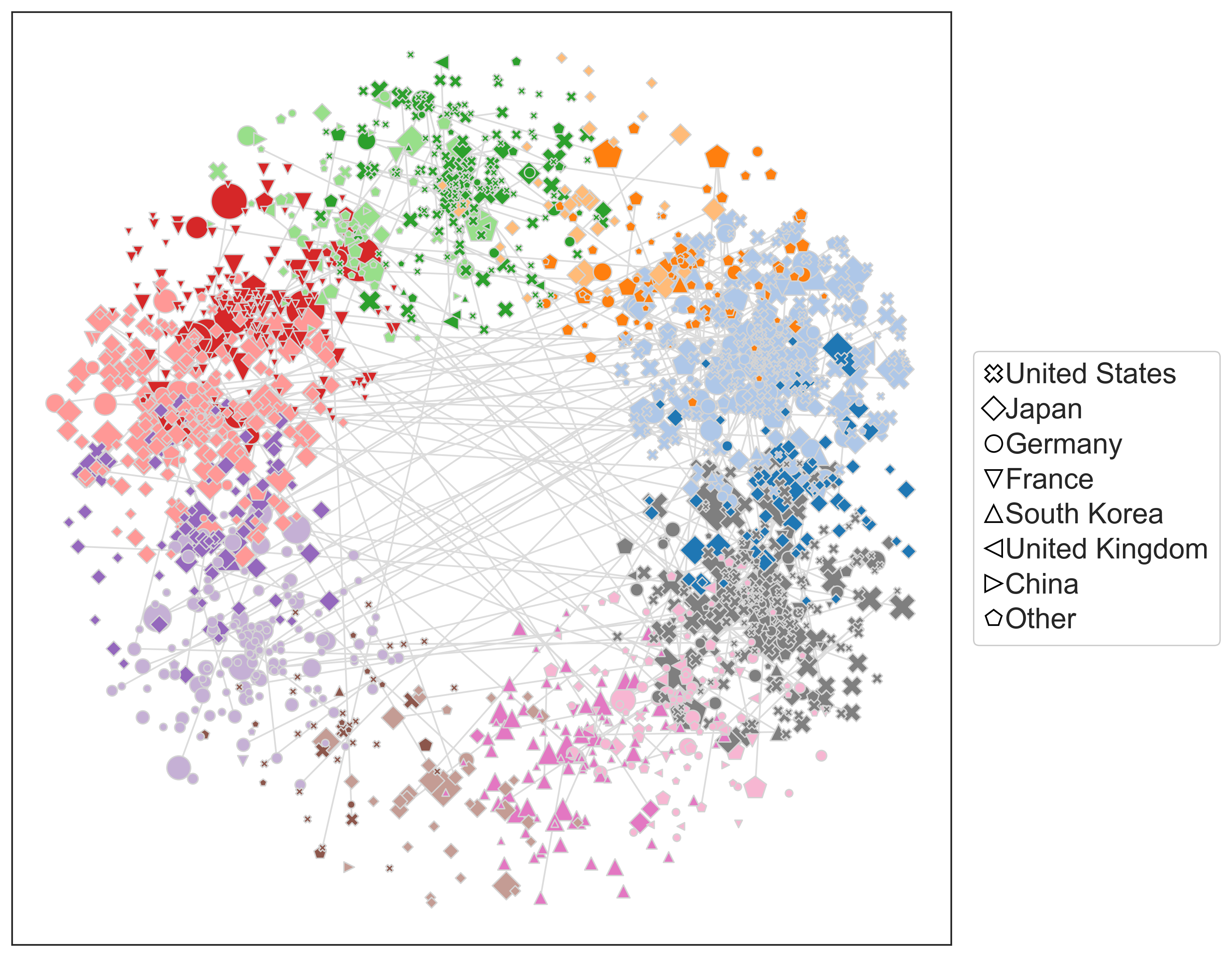}
    \label{fig:communities_net}
\end{figure}

The communities extracted for MULTI are shown for a random subset of patent families in \Cref{fig:communities_net}. Analogous figures for the ALL-AGG and US-AGG networks can be found in \Cref{app:comparison}. While the model allows for overlapping communities (nodes can belong to multiple communities), in \Cref{fig:communities_net} we colour nodes by their `hard' communities, whereby each patent family is assigned to the community to which it displays the highest affinity. The optimal number of communities, calculated via the cross-validation exercise described in \cref{app:CV}, was found to be 15 for the multilayer citation network and 7 for ALL-AGG and US-AGG. Finally, the location of the assignee of each patent family (rather than the priority office, which is used as a covariate in the community detection procedure) is indicated by the shape of the node.

Between networks, several graphical observations can be made in the geographic composition of the extracted communities, despite the differing community sizes. First, country-based homophily is very clear. The most obvious example of this is that families filed by Japan-based assignees are primarily grouped with families that are also filed by Japan-based assignees, with the only observable difference between the networks being how many communities are found within this group of families (1 for ALL-AGG and US-AGG, and 5 for MULTI); however, this difference is expected as the optimal number of communities multilayer network is greater. The other consistently geographically-homogeneous communities include those families assigned to German firms and those assigned to South Korean firms. The existence of these groupings is somewhat expected --- geographic citation biases are a well-known phenomenon and have a wide range of drivers, including local industry agglomeration, shared language, prior-art search strategies, knowledge spillovers, and coordinated technological development strategies at the national level~\citep{jaffe1993geographic,almeida1999localization,macgarvie2005determinants,bacchiocchi2010international,wada2016obstacles}. Because priority office information is included in the community detection algorithm, the existence of the kind of geographic grouping we observe reflects that while technological similarity plays a big role in citation linkage at the micro-level, simple geographical metadata can be highly predictive of network structure at larger scales.

\subsection{Network Communities and Technological Similarity}\label{sec:similarity}

One would expect that the citations we consider in this work \textit{should} link families with technological similarities and, therefore, the communities detected should group inventions with shared and legally relevant technological features. Indeed, the geographical biases in citation linkages that are observed above may be considered to be artifacts of the systems within which technological development occurs, and perhaps even hinder our understanding of the nature of innovation more generally. We assert that the multilayer framework is one way of mitigating some of these biases, as it integrates relevant technological relationships uncovered by several different, and geographically separated, patent agents and examiners working mostly independently. In aggregate, this information should give a more balanced view of technological similarity and down-weight those links that are heavily influenced by unwanted geographical and office-specific biases and conventions. However, the link weights in the ALL-AGG network may play a similar role. As such, we will now turn to the differences in the technological information contained in the three networks and examine the importance of citation context (i.e., source and justification) in assessments of technological similarity.

To do this, we directly compare the network of meso-level technological relationships that can be gleaned from extracted communities with externally-defined technological categories. First, we construct a weighted bipartite (two-mode) network of relationships between the extracted communities and the 3-digit Cooperative Patent Classification (CPC) codes that the families within each community were assigned upon application to the USPTO.\footnote{This choice was made for the sake of consistency. Different offices may make slightly different judgements regarding the particular set of classes assigned to an application. By using data from a single office, we do not have to be concerned with these systematic differences.} CPC codes, henceforth referred to simply as \textit{classes}, were chosen due to their status as the primary classification system at two of the triadic offices and widespread use in research, particularly in studies of technological evolution and forecasting technical change. The weight of each link in the bipartite network between communities and classes is proportional to the fraction of families in each community that were assigned to a given class. We then project onto the technology class nodes to obtain a network of classes wherein links exist between classes that were both found in the same community or communities. A higher link weight between two nodes in this projected network reflects a more similar distribution of those classes across the extracted communities~\citep{vasques2018degree}.\footnote{Note that these classes are not directly used in the community detection process. However, the community detection process relies on citation linkages, and these citations are often found through searches within the technology classes to which the application under examination has been assigned~\citep[see, e.g.,][]{demey2020search}.} 

We then construct two basic comparison networks: co-classification~\citep{engelsman1994patent,breschi2003knowledge} and inter-class citation linkage~\citep{leten2007technological,alstott2017mapping}. The former contains a link between (3-digit CPC) classes when a family is assigned to both, with weights proportional to the relative frequency of such occurrences. The latter network contains links between these classes with weights proportional to the number of citations made between families that were assigned to each class, normalised to the total made by each of the classes.\footnote{To compare this network to our (undirected) projected networks, we take the sum of the normalised weights of the directed links between classes to obtain an undirected link weight. This simplification is a necessary evil for the current purpose and may miss some nuance in certain technological relationships.} We keep self-loops in this network, as they are required for sensible link-weight normalisation. For example, if class A makes 10 citations (and receives none), one of which goes to a class B family but 9 return to other class A families, this is a very different situation from one in which all 10 go to class B families. Because we normalise link weights by total citations made, ignoring self-citations would give the link from A to B the same weight in both scenarios, rather differing by a factor of 10.  Further description of the construction of all networks used in this section can be found in \cref{app:comparison}.

Now that we have two externally defined, node-aligned class networks, we are able to directly compare their structure to those extracted from the community-class bipartite networks. Because the nodes in each network we wish to compare are labelled and the same for all networks, we are able to use known-node correspondence methods that allow for comparisons at the node-level in such a way that accounts for differences in relationships between specific node pairs and for higher-order relationships~\citep{tantardini2019comparing}. For this exercise, we use two different methods of comparison: the Frobenius norm and DeltaCon~\citep{koutra2013deltacon}.

The Frobenius norm is applied to the raw differences in the adjacency matrices between two networks, and thus quantifies the entry-wise (link-level) differences in the matrices being compared. When the networks being compared are unweighted, this distance is simply the square root of the number of pair-wise differences between the networks. However, this method easily accommodates the weighted case, wherein each pair-wise difference can have a magnitude other than unity.\footnote{Specifically, the Frobenius norm of a matrix $A_{m \times n}$ is  defined as the square root of the sum of the absolute squares of its elements,
\begin{equation}
||A||_F = \sqrt{\sum_{i=1}^{m}\sum_{j=1}^{n} |a_{ij}|^2} \, .
\end{equation} } The Frobenius norm is a crude comparison method that cannot account for higher-order relationships between nodes, such as the importance of a link in the overall structure of the network, but it is a good heuristic when making multiple comparisons as we do here. DeltaCon, on the other hand, is more sophisticated, and indirectly considers every possible path between two nodes. In this way, differences in the weights of links that are particularly important for the network structure at the macro-level are incorporated into the comparison. While the DeltaCon algorithm can be very computationally expensive on large networks, and an approximation is possible, our class network is small enough (535 nodes) that the exact form can be used~\citep{koutra2013deltacon}. Both the Frobenius norm and DeltaCon calculate a distance metric whereby smaller distances indicate more similar networks. These methods are implemented in Python using the \url{numpy}~\citep{oliphant2006guide} and \url{netrd} packages~\citep{mccabe2021netrd}.

In addition to the network comparison methods, we are also able to quantify the diversity of technology classes within each community extracted. For this purpose, we make use of the Rao-Stirling diversity (RSD)~\citep{rao1982diversity,stirling2007general}, which considers both the homogeneity of each community (with respect to the classes within it) and the level of `surprise' that specific pairs of classes are found together. For the latter consideration, we operationalise class distance using the inter-class citation network described above, as citations are what we use to extract the communities in the first place.\footnote{That is, it would be a `surprise' to find a pair of classes that don't cite each other, but are nonetheless found in the same community.} Calculating this index for all communities extracted from a particular network, we take the median index across these communities as a measure of their average diversity. The RSD is high for a particular community when classes co-occur in high proportions with other classes with which not many citations are exchanged. RSD is low when classes generally only co-occur in high proportions with other classes with which they exchange many citations. Specifics of the RSD can be found in \Cref{app:diversity}. This kind of analysis, when compared to the network comparison methods above, may be considered relatively myopic. It can only capture the internal composition of individual communities without accounting for the relationships between the pairs of technologies in other communities. However, this calculation may provide insight into the origin of differences we find for the network-level comparisons.

Lastly, we calculate the spread of technologies across the extracted communities. A priori, we do not know what the relationships between the communities are, so we cannot integrate a distance metric to account for the level of surprise that a family assigned a particular class is found in a given pair of communities (as we did for the previous diversity measure). As such, we implement the Herfindahl–Hirschman Index\footnote{Sometimes referred to as the Simpson index.} (HHI)~\citep{hirschman1945national,simpson1949measurement,herfindahl1950concentration,hirschman1964paternity} to measure the extent to which classes are splintered across communities. The details of this calculation can be also found in \Cref{app:diversity}. The HHI is maximised when all families assigned a particular class are in the same community and minimised when there are the same number of these families in each community. Again, the median HHI across all communities is compared across the networks. Like the Rao-Stirling index above, this calculation may add additional colour to the more comprehensive network comparisons. It is important to note that while we believe that it is desirable that communities are able to capture, to some extent, the large-scale structure of the technology-level networks, neither the spread of technologies across communities, nor the internal diversity of communities, is a test of the performance of the community extraction exercise.

It is important to note that the optimal parameters for the community partitions for the three networks are different---the optimal number of communities found for the multilayer network is 15, while for the others it is 7. For this reason, we run the community-detection algorithm for each of the non-optimal partitions (7 for the multilayer network and 15 for the others) to obtain a complete set of networks with which we can make fair comparisons. In sum, we construct six bipartite (community-class) networks which we project onto the class nodes to compare with the co-classification and inter-class citation networks.

\begin{table}[t]
\caption{ \textbf{Network comparison and diversity measures.} Here we show the results of the network distance calculations as well as the diversity measures. DeltaCon (DC) and Frobenius (F) distances between the class-projected networks (leftmost block) and the externally defined co-classification (Co-class) and inter-class citation linkage (IC Cites) networks are displayed in the central block. The median Rao-Stirling class diversity (RSD) across communities and the median Herfindahl–Hirschman Index (HHI) of classes' dispersion across communities are shown in the rightmost block. * indicates non-optimal partition. The lowest values within each comparison set are highlighted in bold. \label{tab:network_metrics} }
\vspace{3mm}
\centering
\renewcommand{\arraystretch}{1.3}

\begin{tabular}{cc|cccc|cc}
& & \multicolumn{4}{c|}{Comparison Network}           & \multicolumn{2}{c}{Diversity} \\ \hline
& & \multicolumn{2}{c}{Co-class} & \multicolumn{2}{c|}{IC Cites} & \multirow{2}{*}{RSD} & \multirow{2}{*}{HHI} \\
& \textit{\textbf{C}} &     DC      &     F     &     DC      &     F     &           &                   \\ \hline
MULTI*  & 7 &   \textbf{32.43}        &  \textbf{115.98}        &   \textbf{30.40}        &    \textbf{115.01}      &   \textbf{6.81}     &     \textbf{0.185}    \\
ALL-AGG  & 7 &    34.00       &   149.40       &   31.96        &    148.31      &      7.02      &   0.263      \\
US-AGG     & 7 &    34.54       &    150.63      &  32.49         &   149.59       &     7.05      &     0.271       \\ \hdashline
MULTI  & 15 &    \textbf{30.08}      &    \textbf{58.66}      &     \textbf{28.09}      &  \textbf{58.61}        &    \textbf{6.87}    &     \textbf{0.131}       \\
ALL-AGG*  & 15 &    31.47       &   78.87       &   29.49        &   78.56       &       \textbf{6.87}    &    0.169       \\
US-AGG*  & 15 &   30.79        &  76.73        &    28.81       &  76.42        &       6.99    &   0.167

\end{tabular}

\end{table}

The results of this analysis can be found in \Cref{tab:network_metrics}. First, we find that the communities in the multilayer network generate class networks that are more similar to the co-classification and inter-class citation networks than those generated by the other two networks. This finding holds for both individual-link-level comparisons (Frobenius) and when higher-order relationships are taken into account (DeltaCon), for both optimal and non-optimal partitions of the multilayer network. Further, we find that the average RSD of the individual communities is lowest, while classes are the most evenly distributed across the communities (low HHI), in the multilayer case.

These observations lend themselves to some interesting interpretations. When looking at all communities, in combination, those extracted from the multilayer network imply technological relationships that are closer to the explicit technology networks than the flattened or single-jurisdiction approaches. However, the diversity calculations suggest that this observation is not simply driven by the extraction of homogeneous communities that group technologies in a straightforward manner. In fact, technology classes are more thinly spread across communities in the multilayer case, while the average internal diversity of classes is generally lowest for this network once known technological similarities between classes are accounted for. This suggests that, on the micro-level, the multilayer (relative to the single-layer) network approach is more sensitive to citation linkages than co-classification, but is nonetheless better able to represent real technological relationships on the meso- and macro-levels.

Indeed, our results are consistent with the conclusion that the multifaceted nature of the technological relationships that are embedded in citation data may be partially lost when a multilayer network is flattened into a single-layer one. This view rests on an assumption that different technology types can be related to each other in different ways. For example, let's assume that applicants filing a patent assigned to class A prefer to cite families assigned to class B, while examiners examining the same patent prefer to cite those assigned to class C. When, such as in this example, these different relationships are driven by different citing parties, the erasure of citation context will lead to the loss of this nuance. This problem may be exacerbated in the presence of higher-order effects, such as if the above citation behaviour only occurs when a fourth class D is also assigned to the patent application. In contrast, the multilayer network approach ensures these nuances and higher-order relationships remain accessible. The retention of this kind of technologically relevant information, particularly with respect to rare or subtle inter-class relationships, would be consistent with the findings displayed in \Cref{tab:network_metrics}.

\section{Discussion and Conclusions}\label{sec:conclusions}

Historically, research informed by patent citation data has often ignored citation source and context. There can be a perfectly reasonable reason for this practice, such as when one is only interested in citations made to and from patents in a single jurisdiction to study, for example, the effect of a local policy change. However, a truly comprehensive and global view of patented inventions and the relationships between them is only possible when data from multiple sources are integrated sensibly. It is in these contexts that the multilayer network is a natural framework for analysis.

In this work, we introduce the concept of multilayer patent citation networks as a natural way to present and analyse global patent information without loss of citation context. We conduct several empirical analyses to demonstrate the utility of the multilayer framework. All analyses are conducted on a subset of the full citation network, containing all triadic patent families classified into CPC class Y02 with US members granted from the year 2001. By design, this subset will give the most conservative estimates of the additional information that may be extracted from the multilayer network relative to its single-layer counterparts. Our results in this work suggest that not only is there, indeed, a considerable amount of additional information contained in the multilayer citation network relative to those single-layer counterparts, but this information is technologically relevant and captures nuanced aspects of the technological relationships between patented inventions.

First, an interdependence analysis shows that additional network layers, defined by citing office, contain complementary (rather than redundant) information that may be used to predict the link-level structure of other layers. To test whether this complementary information is important for characterising network structure more generally, we then conduct an exercise in community detection. This is carried out and compared across three different networks: the multilayer network, the flattened and weighted (single-layer) version of the multilayer network (containing all the links in the latter but without citation context), and the complete (flattened, single-layer) US citation network that is most commonly used in technological network analyses. While there is a notable similarity in the communities extracted from these networks, there is also significant disagreement, indicating that the information contained in the citation context may be important for characterising the mesoscopic structure of the global citation network.

To test whether the differences in community structure are technologically meaningful, we conduct direct comparisons between the technological relationships implied by the extracted communities and those of previously studied meso-scale networks of technological similarity: the co-classification and inter-class citations networks, at the CPC 3-digit level. These tests are conducted, in part, to show how the information content (i.e., citation context) contained in citation networks can be related to the meso-scale technological structures that are perhaps more established in the technology management community. To be able to draw a direct comparison, we construct the bipartite networks between communities and classes, then project onto the class nodes to obtain a class network wherein links reflect levels of co-occurrence in the communities. To add colour to these comparisons, we also compute the Rao-Stirling diversities of these communities (across classes) and the Herfindahl–Hirschman Indices of class (across communities). Relative to the flattened networks, we find that while the communities extracted from the multilayer network are less diverse and the implied class network more similar to the co-classification and inter-class citation networks, classes are more evenly spread across communities. These results suggest that citation context is technologically relevant and a more realistic mesoscopic network structure can be inferred when we depart from the view that technological relationships are mono-faceted or driven by simple class-level technological similarity.

While we include the US citation network in our comparison exercises, this is only done as an acknowledgement of its position as the dominant data source in the extant literature. The flattened version of the multilayer network, on the other hand, contains all the links that are present in the multilayer network, but without the context that allows us to define the layers. As such, we consider this network the most appropriate comparison network, as any differences found must be driven by the absence of citation context. That the communities extracted from the multilayer network more closely replicate the established and explicit co-classification and inter-class citation networks indicates that citation context adds technologically relevant information in the aggregate, despite displaying higher within-community diversity of classes. This suggests that ignoring citation context results in a bias towards within-class citations (that are easier for all parties to search for and find), at the expense of the rarer inter-class citations and class combinations that play a larger role in both the network structure as a whole and, arguably, technological progress in the long-term~\citep{castaldi2015related,verhoeven2016measuring,mewes2019scaling,kelly2021measuring}. Considering citation generation mechanisms, it is plausible that citation context provides important clues as to the relevance and nature of the technological relationship between citing and cited inventions~\citep{criscuolo2008does,alcacer2009applicant,azagra2011smoothing,li2014patent,kuhn2020patent}. As such, treating all these links as equal, with respect to their information content, is clearly not ideal for many use-cases.

\subsection{Limitations}\label{sec:limitations}

The main limitations of the empirical analyses conducted in this work are those restrictions we placed on the families we chose to include. As we describe in \Cref{sec:data}, these restrictions were put in place for a variety of reasons, including data availability, computational limitations, and a desire to demonstrate our approach in a conservative manner. Little can be done about data availability; however, this only affects our ability to examine citation context in the US case, and only for times earlier than the year 2001. In any case, we suggest that families granted after this time provide a sufficiently large sample for the purposes of this work. 

The conservativeness of our approach is introduced with the decision to consider only those families with granted patents at all three triadic offices. This means that all offices had access to the same set of prior art, and had the opportunity to share information among themselves. In turn, this would introduce maximum redundancy between layers, and minimise the additional information that can be added by the inclusion of citation context. It is for this reason that we think of our approach as conservative. Extensions of the restrictive, special-case multilayer framework that we examine here are discussed below in \Cref{sec:future}, and highlight the potential of this framework going forward.

Lastly, to reduce the computational complexity of our analyses, we restrict the included families to those classified into CPC class Y02. While we maintain that this subset is an appropriate representation of the patent citation network as a whole, there may be arguments against its generalisability. However, in the case that this class contains a more homogeneous set of families than the set of all families (which is almost certainly true), then the inter-class structure that we are able to explore is likely to be \textit{less} rich and \textit{less} nuanced than that of the full network. Detecting higher-order nuances is precisely the domain in which we suggest the multilayer network excels, so following this logic would lead us to conclude that the current approach is, again, a very conservative one.

\subsection{Future Work}\label{sec:future}

This work aims to describe the construction of multilayer patent citation networks then conceptually and empirically justify their use. This framework may prove to be of particular interest to those who would prefer representations of technological relationships that are not as sensitive as extant frameworks to the idiosyncrasies of individual patent offices. However, both the layers that are selected to comprise the network and the appropriate empirical methods to extract information from this network will depend on the specific use-case. Here, we describe the myriad methodological doors that are opened with the introduction of patent-based multilayer networks into the broad field of science, technological, and innovation studies.

The obvious extension to the current work is to take a less conservative approach with respect to the subset of families and citations considered. This can take the form of additional layers, nodes, or links. The addition of layers corresponds to the addition of new citation contexts (such as in-text citations~\citep{verluise2020missing}) or the addition of new jurisdictions. The addition of nodes and links, on the other hand, would relax the condition that a family be triadic. Citations between triadic families only make up a tiny portion of all citations made and received by these families.  For example, in \Cref{fig:multilayer}, we show the triadic ego network of the family with USPTO equivalent US-6819081-B2. In this restricted network, this family only receives 4 citations from other triadic families classified into class Y02. If we remove all restrictions on the patents we include in our network, however, this family receives almost 50 citations; about 90\% of these are from families that have a triadic member, and about 95\%  are from families that are also classified into Y02. As such, removing the triadic family requirement but keeping the network restricted to the triadic offices and the Y02 classification would dramatically increase the sample size.

Multilayer citation networks can also be flexibly aggregated. Just as one can analyse the inter-class citation network for a single jurisdiction or citation context (e.g., US applicant citations), it is also possible to include additional layers containing the equivalent information for other jurisdictions or contexts. In fact, in the same way that we use families to align layers in the current work, any metadata that connects groups of patents between network layers forms a natural multilayer configuration. Classes, firms, and inventors can all be linked across jurisdictions and citation contexts and their networks analysed in a multilayer framework. Even in the single-jurisdiction US case, for example, the relative positions of firms in the inter-firm citation network will depend on whether one uses examiner citations, applicant citations, third-party citations, or in-text citations. Because firms can be represented as nodes across all of these context-specific networks, multilayer network tools may be applied to obtain a comprehensive and integrative view of the network structure without abandoning citation context. Citations to non-patent literature such as scientific articles is challenging to incorporate into patent citations networks generally, but it is certainly possible to treat this information as family-level metadata --- perhaps to construct a bipartite network similarly to how technology classification was used in \cref{sec:similarity}. More complicated uses of this information could match institution and inventor data from patents onto scientific articles to extend recent work on the multilayered interplay between authorship and the broader dynamics of science and collaboration into the technological domain~\citep{omodei2017evaluating,nanumyan2020multilayer,zingg2020citations}.

In addition to these data extensions, the conceptual arguments against the omission of citation context lead to a strong case for the further application of novel tools designed for the study of multilayered systems. To return to the public transport analogy, it would be unwise to treat all modes of transport as equal if you are trying to find the fastest route between two places in the network. In the same way that the time and financial costs of using different modes affects the route choice between two points in a physical landscape (which will be moderated by the amount of time or money you had), citation networks are embedded in a technological landscape~\citep{kauffman2000optimal,fleming2001technology} and different types of citation may traverse this landscape in different ways. This intuition has significant consequences for the analysis of citation networks. For example, any algorithm that `walks' through the network, such as PageRank, should consider the `cost' of each link in a similar way to one plotting a route through a multilayered transportation network. 
The application of multilayer network methods opens the door to a menagerie of new analytical tools to develop more sophisticated and tailored metrics for studies of technical change and the nature of innovation systems. For example, the identification of patent thickets~\citep{shapiro2000navigating,bessen2003patent} is often conducted through, or supported by, citation network analysis~\citep{von2011measure,zingg2018nanotechnology,yuan2020network}. The multilayer framework may assist in these studies --- thicket identification depends crucially on the citation context (blocking vs. non-blocking citations) and the jurisdiction (a thicket is necessarily a single-jurisdiction phenomenon). Adding citation context and linking families across jurisdictions for direct comparison may allow for thickets to be more easily distinguished from fields with dense, but non-overlapping, intellectual property rights. For example, when calculating clustering coefficients in multilayer networks, one can specify weights for different kinds of citation or penalise cycles that move between layers~\citep{de2013mathematical}. This kind of flexibility can be used to operationalise the definition of thickets in a way that doesn't simply ignore applicant-provided citations or citations from other jurisdictions, which may not be entirely irrelevant, particularly at the firm level.

Network centrality is another important concept that is generalised in the multilayer case~\citep{sola2013eigenvector,sole2014centrality,de2015ranking,taylor2021tunable}, and can also be readily applied to citation networks. For example, without citation context, it is hard to know whether firms are central because they block the patents of competitors or are a source of knowledge from which other firms build. Further, firm centrality will likely depend on the jurisdiction one examines, so multilayer centrality may give a more holistic view of their centrality in global markets.

Both technology roadmaps~\citep{lee2009business} and technological trajectories~\citep{verspagen2007mapping} may be significantly altered by the incorporation of citation context, as different kinds of citation appear to hold different information, which may, in turn, be useful for forecasting or tracing different kinds of technical change~\citep{acemoglu2016innovation,mariani2019early}. So-called `main paths' in technological trajectory analysis~\citep{hummon1989connectivity,verspagen2007mapping} could be particularly sensitive to the weights that are placed on, or empirically determined for, different layers or citation contexts. The multilayer framework may also conceptually aid traditional economic analyses~\citep{cai2019growth}, for which it is possible, for example, to allow layers to differ in importance when constructing proxy network variables that attempt to capture an abstract concept.

Lastly, pair-wise interactions may not be sufficient to describe the complex behaviour of interactions between the components of innovation systems that are accessible through citation networks. In particular, the interactions between firms or technology types that are visible in citation networks may be better represented through higher-order interactions~\citep{lambiotte2019networks,battiston2020networks,battiston2021physics}. For example, the patenting and citing behaviour of firms may be described at several different scales. Higher-order representations allow us to differentiate changes in citation behaviour of a firm in response to sector-wide changes from the pairwise interactions between a firm and every other firm in its sector. Higher-order interactions can exist within layers of multilayer networks and it is possible that different higher-order behaviours are observable in different patent systems. In any case, it is clear that applications of network frameworks beyond single-layer networks with dyadic links are very much in their infancy in the field of innovation studies, and hold huge potential as more realistic abstractions of innovation systems.

\subsection*{Acknowledgements}
The authors thank the International Max Planck Research School for Intelligent Systems (IMPRS-IS) for supporting Martina Contisciani. MC and CDB were supported by the Cyber Valley Research Fund.

\clearpage

\bibliography{main}

\clearpage
\appendix
\appendixpage
\pagenumbering{roman}
\section{Model description}\label{app:MTCOV}

For the layer interdependence and community detection analysis we use MTCOV,\footnote{\url{https://github.com/mcontisc/MTCOV}} the model developed by \cite{contisciani2020community}. MTCOV is a probabilistic generative model that incorporates both the topology of interactions and node attributes to extract overlapping communities in directed and undirected multilayer networks. It works also with single-layer networks, since this is the special case for which there is only one layer in the `multilayer' network. The model assumes conditional independence between the network and attribute data, given a set of latent variables (including the node community memberships). The likelihood function is a linear combination of the network and attribute information, adjusted by a scaling hyperparameter $\gamma \in [0,1]$, which controls the relative contribution of the two terms: for $\gamma=0$ the model only considers the network topology, while for $\gamma=1$ it only considers the attribute information.

MTCOV has four parameters: two membership matrices accounting for outgoing and incoming links respectively, an affinity tensor that describes the density of links between each pair of groups among the different layers, and a parameter that matches communities and node attributes. The inference is performed with an Expectation-Maximization algorithm, and its implementation is efficient and scales to large datasets (such as the one studied here) because it exploits the sparsity of the dataset.

\subsection{Cross-validation and hyperparameter settings}\label{app:CV}

MTCOV has two hyperparameters, the scaling parameter $\gamma$ and the number of communities $C$. For each network under analysis, we estimate the hyperparameters by using 5-fold cross-validation along with a grid-search to range across their possible values. For the current work, we choose to vary $C \in \{2, 3, 5, 7, 10, 12, 15\}$ and $\gamma \in \{0, 0.3, 0.5, 0.7, 1\}$. Specifically, we divide the dataset into five equal-size groups (folds), selected uniformly at random, and give the models access to four groups (training data) to learn the parameters; this contains 80\% of the matrix entries and covariates. One then predicts both links and node attributes in the held-out group (test set). By varying which group we use as the test set, we get five trials per realization. For performance metrics, we measure the area under the receiver-operator characteristic curve (AUC) (for the link prediction) and the accuracy (for the node attribute prediction) on the test data, and the final results are averages over the five folds. The AUC is the probability that a random true positive is ranked above a random true negative; thus the AUC is 1 for perfect prediction, and 0.5 for random chance. The accuracy classification score is 1 for perfect recovery and 0 in the worst case of overfitting. In order to choose the best pair of hyperparameters $(\hat{C}, \hat{\gamma})$ we look for the pair that performs best across both AUC and accuracy in the test set. 

Since the networks are large, it is not always possible to compute the AUC on the whole training and test sets, hence we proceed with samples. In detail, we fix the number of comparisons we want to evaluate, here $10^5$, and for both the train and the test sets we sample $10^5$ values from zeros entries (where there is no existing link) and we compute the link prediction on that sample (we save these values in a vector $R_0$); we do the same with the non-zeros entries (we save these values in a vector $R_1$). We then make element-wise comparisons and compute the AUC as: 
\begin{equation} 
\label{eq:AUC}
AUC = \frac{\sum{(R_1 > R_0)} + 0.5\sum{(R_1==R_0)}}{|R_1|}
\end{equation} 
where $\sum{(R_1 > R_0)}$ stands for the number of times $R_1$ has a higher value than $R_0$ in the element-wise comparison; and $|R_1|=|R_0|$ is the length of the vector which is equal to the number of comparisons we fix. 
Moreover, when the network has a number of nodes bigger than 5000, we run the algorithm by computing the likelihood only on a batch of nodes (here a random subset with 5000 nodes) to speed up the computational time. 

\Cref{tab:hyperparameters} shows the optimal hyperparameters obtained for all single-layer and multilayer networks used in the manuscript.

\begin{table}[t]
\caption{ \textbf{Hyperparameters setting.} Values of the hyperparameters $C$ and $\gamma$ extracted by 5-fold cross-validation combined with grid-search. \label{tab:hyperparameters} }
\vspace{3mm}
\centering
\renewcommand{\arraystretch}{1.2}

\resizebox{\textwidth}{!}{\begin{tabular}{ccccccccccc}
 & US-EXM & US-APP & EP-APP & EP-ISR & EP-SEA & JP-REJ & JP-BCK & US-AGG & ALL-AGG & MULTI \\ \hline
 $C$ & 7 & 7 & 7 & 7 & 3 & 7 & 7 & 7 & 7 & 15 \\ 
 $\gamma$ & 0.3 & 0.7 & 0.7 & 0.7 & 0.7 & 0.7 & 0.7 & 0.7 & 0.7 & 0.7 \\
\end{tabular}}

\end{table}

\subsection{Layer interdependence analysis}\label{app:interdepedence_calc}

The layer interdependence problem consists of identifying which sets of layers are structurally related, and quantifying the strengths of those relationships. To this end, we use the MTCOV model and we employ the method described in \cite{de2017community}. This method consists of performing link prediction in one layer with and without the information in another layer to quantify the extent to which these two layers are related. Thus, for our purposes, interdependence is based on the idea that two layers are interdependent if the structure of one layer provides meaningful knowledge about the structure of the other. 

To test our ability to predict a set of target layers $\alpha$, we perform experiments with 5-fold cross-validation following the same routine as above by using only the optimal pair of hyperparameters. The main difference from the community-detection procedure above is the way the training and test sets are built. In fact, for the layer interdependence task, we only split (5-fold) the links in the set of target layers $\alpha$ together with the attributes for the nodes in this set, while giving full access to the set of layers $\beta$ when they are added.

For this task, because we are mainly interested in link prediction, rather than in recovering covariates, we measure the AUC as in \Cref{eq:AUC}. The final AUC is the average obtained over the five folds, each of which holds out a different subset of 20\% of $\alpha$. The value of the AUC depends both on the set of target layers $\alpha$ we are trying to predict, and on what set of other layers $\beta$ we give the algorithm access to.

As described in \cref{sec:interdependence}, we restrict our analysis to the sublayers generated by the USPTO (separately) and the JPO and EPO layers (as sets of sublayers), without exploring all possible combinations of sublayers. In detail, we consider the following experiments:
\begin{enumerate}[(a)]
\item $\alpha$ = [US-APP], $\beta_1$ = [US-EXM], and $\beta_2$ = [US-EXM, EPO, JPO].
\item $\alpha$ = [US-EXM], $\beta_1$ = [US-APP], and $\beta_2$ = [US-APP, EPO, JPO].
\item $\alpha$ = [EPO, JPO], $\beta_1$ = [US-APP], $\beta_2$ = [US-EXM], and $\beta_3$ = [US-APP, US-EXM].
\item $\alpha$ = [US-APP, EPO, JPO], and $\beta_1$ = [US-EXM].
\item $\alpha$ = [US-EXM, EPO, JPO], and $\beta_1$ = [US-APP].
\end{enumerate}
Note that for the JPO and EPO, we are using all sublayers of these two jurisdictions. Furthermore, when the set $\alpha$ contains only a sublayer of USPTO [(a), (b)], the hyperparameters used by the algorithm are $C=7$ and $\gamma=0.7$, which is the optimal choice for the US-AGG network. For [(c), (d), (e)] the algorithm uses $C=15$ and $\gamma=0.7$, which is the optimal choice for the multilayer network, for computational simplicity.\footnote{A cross-validation procedure to detect the best pair for the different sets $\alpha$ was determined to be too computationally expensive.}

\section{Network comparison}\label{app:comparison}

\subsection{Class network construction}

\begin{figure}[t]
    \caption{\textbf{Community extraction for comparison networks.} This diagram shows the hard community membership partitions for the ALL-AGG and US-AGG networks. As for \Cref{fig:communities_net}, we use a random sample of 2000 nodes and include any incidental links. The layout is determined by the results for MULTI, for purposes of direct comparison, while the colouring shows the communities found for each network (7 communities for each of ALL-AGG and US-AGG). Node size is proportional to the number of outgoing and incoming citations, while node shapes denote the location of the assignee of each patent family.}
    \vspace{3mm}
    \centering
    \includegraphics[width=\textwidth]{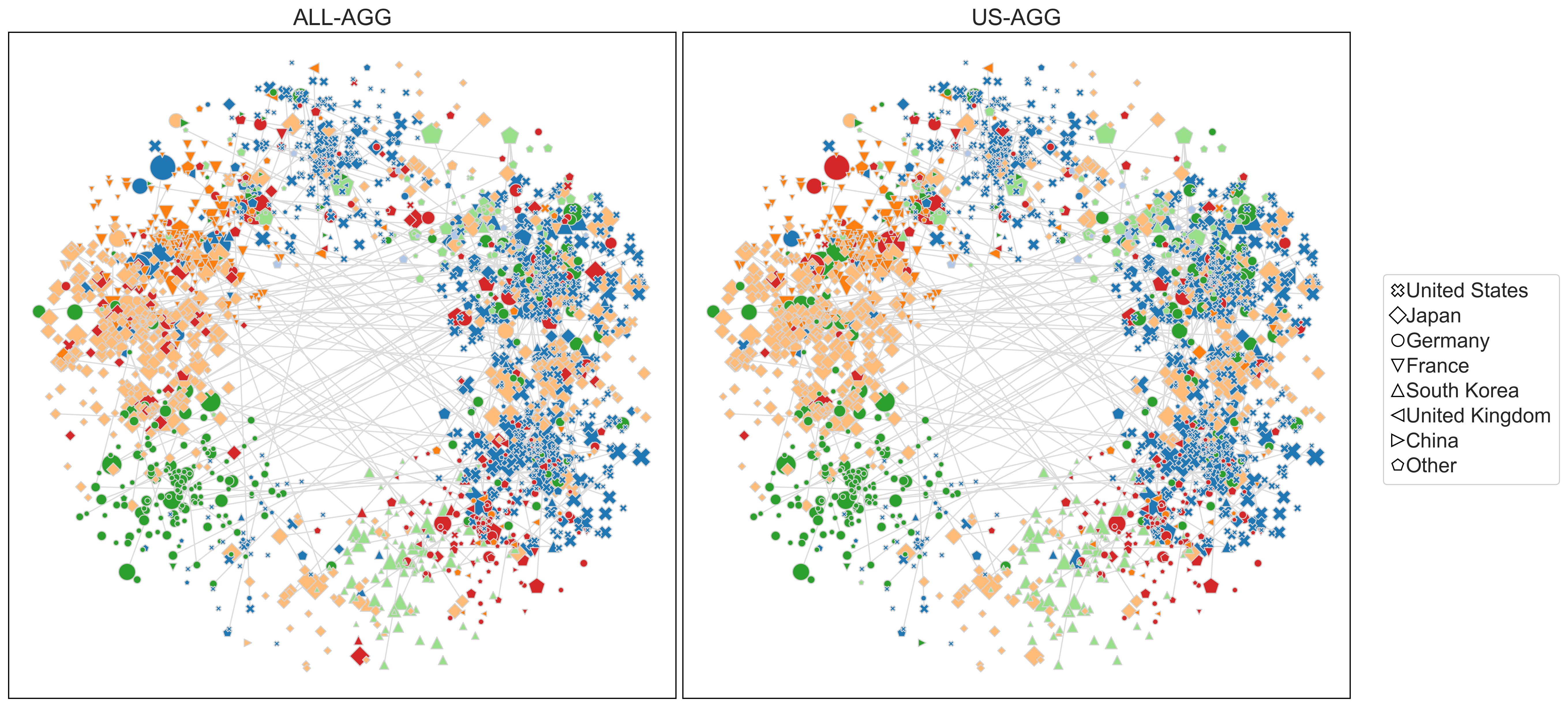}
    \label{fig:communities_net_comparison}
\end{figure}

We use network comparison methods in order to quantify the differences in the technological information contained in the MULTI, ALL-AGG, and US-AGG networks. In particular, we directly compare a projection of the bipartite network of relationships between the extracted communities and the 3-digit Cooperative Patent Classification (CPC) classes with co-classification and inter-class citation networks. \Cref{fig:communities_net_comparison} displays the communities extracted for a random subset of the nodes and edges in ALL-AGG and US-AGG.

To build the bipartite network between communities and classes, we first populate a matrix $P$ whose dimensions are given by the number of families (22653) times the number of classes (535). This is a binary matrix with non-zero entries when a family is assigned to a given class. We then normalize the matrix such that each column sums up to one. In this way, we can consider the matrix $P$ to be the membership matrix of the classes among the patents. By multiplying the transpose of the membership matrix of the patents among the communities and the previous matrix $P$, we get the bipartite network $D = U^{T} \, P$ of relationships between the extracted communities and the classes. 
To ease the comparisons, we need to project this bipartite network onto the technology classes to obtain a network of classes. The projection onto the class nodes is computed through the matrix multiplication $D^{T}\, D$ between the bipartite matrix $D$ and its transpose. This projection has non-zero entries when pairs of classes are both found in the same communities, with weights proportional to their relative frequencies within those communities. As baseline comparisons, we use the co-classification and the inter-class citation networks. The former is obtained by the matrix multiplication $P^T P$, while the latter is constructed as described in \cref{sec:similarity}.

After running the community-detection algorithm for both the optimal and non-optimal partition of each of the three networks MULTI, ALL-AGG, and US-AGG, and only then can we obtain six projected networks among which we are able to make fair comparisons. \Cref{tab:nonoptimal-C} shows the performance of MTCOV on the citation networks (with non-optimal parameters identified with the symbol $*$) for the link prediction (AUC) and covariate prediction (accuracy) tasks, using 5-fold cross-validation.

\begin{table}[t]
\caption{ \textbf{Results of link prediction and covariate prediction tasks}. We measure AUC (link prediction) and accuracy (covariate prediction) over 5-fold cross-validation for $C$ equal to $7$ (the optimal value for ALL-AGG and US-AGG) and $15$ (the optimal value for the MULTI network); $\gamma=0.7$ (the optimal value for all the networks). \label{tab:nonoptimal-C} }
\vspace{3mm}
\centering
\renewcommand{\arraystretch}{1.2}

\begin{tabular}{cc|cc}
& \textit{\textbf{C}} &     AUC      &     Accuracy   \\ \hline
\multirow{2}{*}{MULTI}   & 7*     & 0.835 & 0.341 \\
                         & 15    & 0.852 & 0.422 \\
\multirow{2}{*}{ALL-AGG}    & 7   & 0.730 & 0.402 \\
                         & 15*  & 0.739 & 0.393 \\
\multirow{2}{*}{US-AGG} & 7      & 0.736 & 0.426 \\
                         & 15*     & 0.749 & 0.406

\end{tabular}
\end{table}

After extracting communities, we construct the six bipartite (community-class) networks which we then project onto the class nodes to compare with the co-classification and inter-class citation networks.

\subsection{Diversity measures}\label{app:diversity}

Two diversity measures are used in the main body of this work: Rao-Stirling diversity (RSD) and the Herfindahl–Hirschman Index (HHI). For each network, RSD is calculated at the \textit{extracted-community level} and then a median is taken across communities. The RSD for community $c$ is calculated as~\citep{stirling2007general}:
\begin{align}
    RSD_c=\sum_{i,j,i \neq j} d_{ij} \, p_{i,c} \, p_{j,c} \, ,
\end{align}
\noindent where $d_{ij}$ is a known distance measure between 3-digit CPC technology classes $i$ and $j$, while $p_{i}$ and $p_{j}$ are the proportion of families in the community that are assigned classes $i$ and $j$, respectively. Two factors complicate this calculation. First, because each family can be assigned multiple categories, RSD can take on values greater than one. Because we are directly comparing the RSD for the same set of families (our networks have the same set of nodes), this is not a concern. In fact, we believe this is sensible for this data. That is, if a community consists of a set of families that are all assigned the same two classes $i$ and $j$, our procedure here will treat these communities as consisting of 100\% $i$ and 100\% $j$ (minimal diversity) rather than 50\% $i$ and 50\% $j$ (maximum diversity), for a given $d_{ij}$. Second, because we allow overlapping communities (i.e., a node can be assigned multiple communities with different weights), $p_{i}$ and $p_{j}$ are the weighted sums over patent families $f$ in $c$:
\begin{align}
p_{i,c} = \frac{\sum_{f \in i} w_{f,c}}{\sum_{\forall f} w_{f,c}}  \, , \label{eq:pic}
\end{align}
\noindent where $w_{f,c} \in [0,1]$ is the weight of family $f$ that is assigned to $c$.

For our purposes, $d_{ij}$ is one minus the normalised link weight in the inter-class citation network constructed for our network comparison calculations. This metric is scaled such that distance zero corresponds to the strongest citation linkage for each class, and distance one corresponds to no citation linkage. These new weights act as proxies for the level of surprise, where a weight of zero indicates two classes that only ever cite each other, while a weight of unity indicates two classes that never cite each other. As such, the `level of surprise' parameter $d_{ij}$ down-weights combinations that we expect while exaggerating those that we don't. This adjustment is important. For any given technology class, the number of classes with which it shares community membership depends crucially on both the classification system and the level of the hierarchy within this system that we choose to use. When a class starts to get too crowded, for example, it may be split to make technical search easier~\citep{lafond2019long} --- after all, this is one of the primary goals of patent classification systems. For this reason, a distance measure like $d_{ij}$ is crucial to incorporate into technological diversity measurements.

HHI, also called the Simpson diversity index, is calculated at the \textit{technology level}, $i$, to measure the extent to which technology classes are split across extracted communities. A median across technology classes is then calculated. The HHI for class $i$ is calculated as:
\begin{align} 
    HHI_i=\frac{N\,\sum_{c} p_{i,c}^2-1}{N-1} \, , \label{eq:hhi}
\end{align}

\noindent where $p_{i,c}$ is defined as in \Cref{eq:pic} and $N$ is the total number of communities into which families can be assigned (7 or 15, in our case). \Cref{eq:hhi} is the unbiased version of the HHI~\citep{hall2005note}; this version corrects the $1/N$ offset that affects the standard version of the HHI (for which $1/N$ is the minimum value), which is the sum in the numerator of \Cref{eq:hhi}. The HHI measures how much a technology class is splintered across communities, ranging from HHI=0 for maximally spread to HHI=1 for maximally concentrated. We note that the goal of the community detection process was \textit{not} to replicate the CPC system as closely as possible. There are many valid reasons why a technology class may be split across communities, such as when a technology is particularly generalisable and is applied to (and cited by) many seemingly unrelated fields. Instead, the HHI gives us an idea of what is, or is not, driving the results we obtain for the direct network comparison.

\section{Ego network details}\label{app:ego_details}

The diagram in \Cref{fig:multilayer} shows the multilayer ego network of a triadic patent family, labelled \textbf{A}. \Cref{tab:ego_details} lists the seven families in this diagram, alongside their granted equivalents.

\begin{table}[h!]
\caption{ \textbf{Example network subset details.} Details of each of the families displayed in \Cref{fig:multilayer} are shown below. Priority indicates the month of first filing. All families consist of three triadic patents except for \textbf{C}, which includes multiple family members at the USPTO and EPO. \label{tab:ego_details} }
\centering
\vspace{3mm}
\renewcommand{\arraystretch}{1.2}
\begin{tabular}{c|p{2cm}|p{2cm}p{2cm}p{2cm}|c}
 & \multirow{2}{2cm}{\centering DOCDB Family} & \multicolumn{3}{c}{Equivalent} &  \\
Node &  & USPTO & EPO & JPO & Priority \\ \hline
\textbf{A} & 19192289 & 6819081 & 1333511 & 3671007 & 2002-01 \\
\textbf{B} & 17414436 & 6174618 & 0905803 & 3777748 & 1997-09 \\
\textbf{C} & 26411133 & 6211645, 6211646 & 0892450, 1030389, 1030390 & 4487967 & 1997-03 \\
\textbf{D} & 36242735 & 7615309 & 1695401 & 4527117 & 2003-12 \\
\textbf{E} & 37115311 & 7687192 & 1872418 & 4739405 & 2005-04 \\
\textbf{F} & 38522869 & 7967506 & 1994626 & 5133335 & 2005-03 \\
\textbf{G} & 37115315 & 7488201 & 1872421 & 4663781 & 2005-04
\end{tabular}
\end{table}

\end{document}